\documentclass[aps,prl,reprint,groupedaddress]{revtex4-2}

\usepackage[dvipdfmx]{graphicx}
\usepackage{color}
\usepackage{amsmath}
\usepackage{bm}
\usepackage{comment}
\begin{document}


\title{First-order nuclear dipolar order in rotating solids}


\author{Kohei Suzuki}
\affiliation{Division of Chemistry, Graduate School of Science, Kyoto University}

\author{Kazuyuki Takeda}
\email{takezo@kuchem.kyoto-u.ac.jp}
\affiliation{Division of Chemistry, Graduate School of Science, Kyoto University}


\date{\today}

\begin{abstract}
Nuclear spins' dipolar order is created under magic angle spinning through the first-order process made possible by simultaneous implementation of dipolar recoupling and adiabatic demagnetization in a reference frame reached out through nested transformations, firstly from the laboratory frame into the rotating frame, and then into the spin coordinate system nutating inside its parent frame. In such a nutating frame, both the static and resonantly rotating radio-frequency (RF) fields are invisible, and the re-introduced dipolar interaction provides a secular eigenstate on which dipolar order develops with assist of an additional portion of RF field designed to implement adiabatic demagnetization in the nutating frame.
\end{abstract}


\maketitle

Reversible allocation of a quantum system's populations between eigenstates of one interaction Hamiltonian and those of another provide a basis for quantum control, storage, and measurements, also fostering fundamental concepts in physics.
Interconversion between Zeeman order and dipolar order realized through adiabatic demagnetization and remagnetization\cite{Pound1951,Purcell1951,Ramsey1951} in dipolar coupled nuclear spins in solids led to the concept of spin temperature\cite{Casimir1938,Purcell1951,Redfield1955,Abragam1958,Lurie1964,Jeener1965,Redfield1969} and enabled NMR characterization of superconductors\cite{Hebel1959,Redfield1959}.
Dipolar order can also develop in a persistent static magnetic field by adiabatic demagnetization in the rotating frame (ADRF)\cite{Slichter1961,Anderson1962} or the Jeener-Broekaert (JB) scheme\cite{Jeener1964,Jeener1965,Jeener1967}.
Rotating-frame dipolar order of abundant spins, such as $^{7}$Li, $^{1}$H, and $^{19}$F, can be transported to dilute spin species, like $^{6}$Li, $^{13}$C, and $^{43}$Ca, in the form of an enhanced magnetization of the latter, thereby improving the sensitivity of solid-state NMR\cite{Lurie1964,Freeman1965,McArthur1969,Demco1975}.
Dipolar-order-mediated polarization transfer initiated a number of NMR researches in \textit{stationary} solids including its combination with dynamic nuclear polarization and dissolution NMR\cite{Elliott2020,Elliott2021a,Elliott2021b}.
Thus, dipolar order has significance in terms of both fundamental and practical perspectives.


However, dipolar order has been incompatible with magic angle spinning (MAS)\cite{Andrew1958,Lowe1959,Andrew1959,Maricq1979}, a vital pillar of \textit{high-resolution} solid-state NMR\cite{Mehring1983}.
MAS averages out dipolar interactions, namely, the very interactions that the dipolar order develops upon. 
So far, creation of dipolar order under MAS has been a \textit{second-order} process\cite{Charpentier2002,Ohashi2007,VanBeek2011,Wolf2023}, because the relevant dipolar interaction can arise only through the second-order contribution of at least three coupled spins to the average Hamiltonian\cite{Haeberlen1968}.

Here, we report creation of \textit{first-order} dipolar order under MAS without relying on the second-order effects.
First-order processes are driven by the secular, first-order average Hamiltonian\cite{Haeberlen1968}, so, unlike the previous second-order operations, a pair of dipolar-coupled spins suffice.
The key idea is simultaneous implementation of dipolar recoupling\cite{Oas1988,Raleigh1988,Levitt2007} and dipolar-order creation.
Even though these two have been thought of being incompatible with each other in the conventional sense, we show that they go well when moving into an interaction frame with respect to the radio-frequency (RF) Zeeman interaction that drives continuous spin nutation.
In this what is called the \textit{nutating frame}\cite{Grzesiek1995}, the nutation RF field is hidden, just as the static field in the laboratory frame disappears in the rotating frame\cite{Abragam}. 
By applying an additional portion of time dependent RF field that serves for adiabatic demagnetization in the nutating frame, dubbed here as ADNF, a nutating-frame version of dipolar order can develop\cite{Kunitomo1971}.
Unlike ADRF where the RF field is ramped off, the persistent RF field can be set to cause dipolar recoupling throughout the processes of ADNF, dipolar-order retention, and adiabatic \textit{remagnetization} in the nutating frame (ARNF).
While dipolar recoupling has so far been an obstacle, we \textit{capitalize on} it, providing the spins with a new platform for dipolar order to develop on and making dipolar order compatible with MAS.

\begin{figure*}[htbp]
    \centering
    \includegraphics[width=0.8\linewidth]{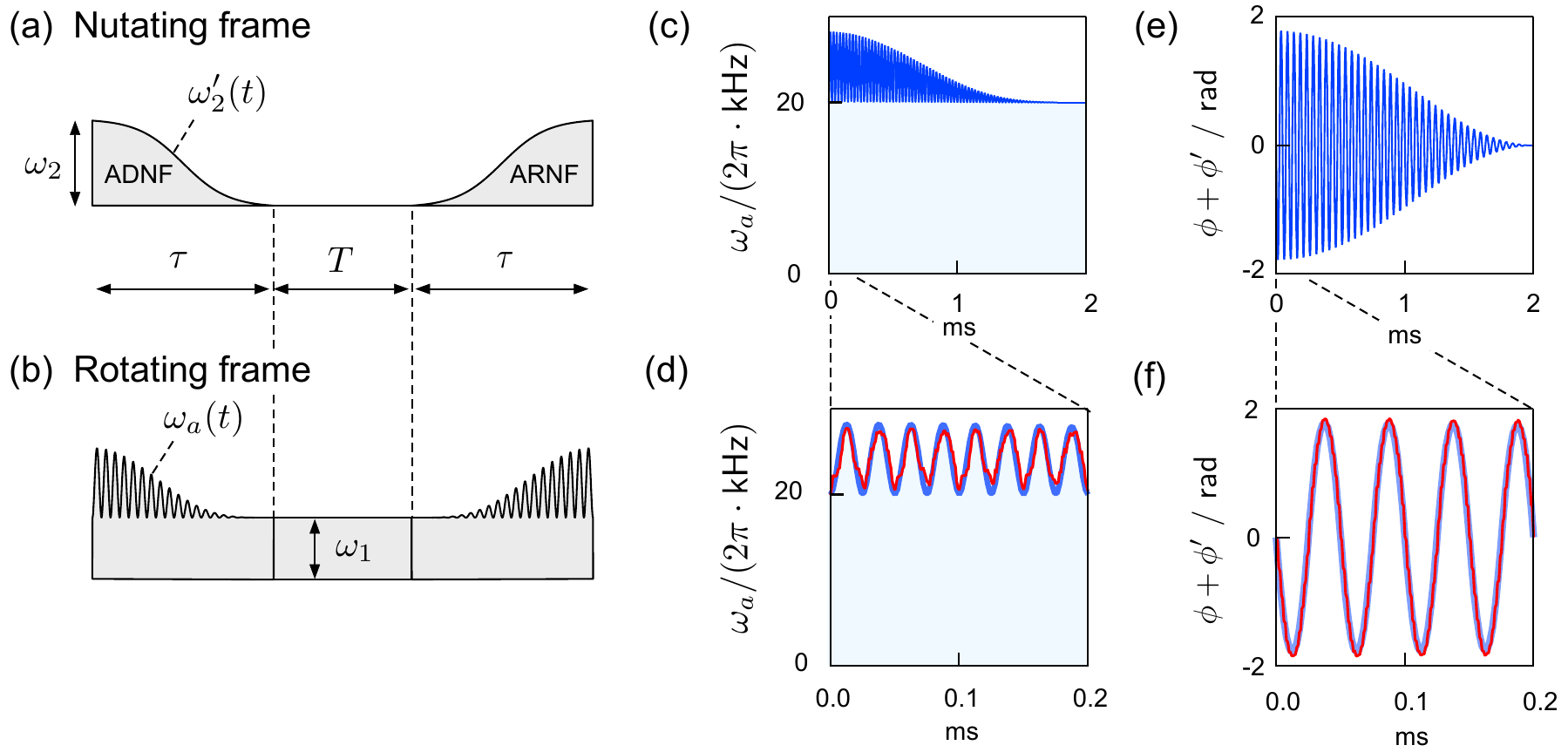}
    \caption{\label{fig:rf-profile} Schematically drawn profiles of (a) $\omega_{2}'(t)$ and (b) $\omega_{a}(t)$ in ADNF followed by dipolar-order retention and ARNF. Fow $\omega_{1}=\omega_{2}=2\pi\cdot 20$~kHz and $\tau=2$~ms, RF amplitude $\omega_{\mathrm{a}}$ and phase $\phi'$ are plotted in (c) and (e) according to Eqs.~(\ref{eq:omega_a})-(\ref{eq:phiprime}). (d) and (f) are magnification of (c) and (d) for a shorter time interval of 0.2 ms, which also show experimental RF-field profile measured through a loopback measurements, in which the RF signal at a carrier frequency of 400.2~MHz going out of the transmitter was sent back to the receiver and quadrature demodulated.}
\end{figure*}
%

First-order dipolar order is creatable in the minimum system of a pair of dipolar-coupled spins, $\bm{I}_{1}$ and $\bm{I}_{2}$.
We consider a static magnetic field $\bm{B}_{0} \parallel z$ much larger than the local dipolar field, a continuous, resonantly rotating RF field $\bm{B}_{1}$, and in addition, a residual, time-dependent RF-field component, denoted here as $\bm{B}_{2}(t)$, which is to be modulated strategically. 
With the RF Zeeman interactions $\mathcal{H}_{1}= -\gamma \bm{I} \cdot \bm{B}_{1} = \omega_{1} I_{x}$~($\omega_{1} \equiv -\gamma B_{1}$) and $\mathcal{H}_{2}= -\gamma \bm{I} \cdot \bm{B}_{2}$, where $\bm{I}=\bm{I}_{1} + \bm{I}_{2}$ and $\gamma$ is the gyromagnetic ratio, time evolution of the spin system in the rotating frame is driven by a propagator $U = T \exp [-i \int dt' (\mathcal{H}_{1} + \mathcal{H}_{2} )]$, where $T$ is a Dyson time-ordering operator.
$U$ can be expressed as a product of propagators $U = U_{1} U_{2}$\cite{Mehring1983}, with $U_{1} = \exp\left[-i \omega_{1} t I_{x} \right]$ and
$U_{2} = T \exp\left[-i\int dt' \mathcal{H}_{2}^{\mathrm{nut}} \right]$ where $ \mathcal{H}_{2}^{\mathrm{nut}}= U_{1}^{-1} \mathcal{H}_{2} U_{1}$.

$U_{2}$ represents the target propagator in the nutating frame.
Setting
\begin{align}
  U_{2} = \exp\left[ -i \int dt' \omega'_{2}(t') 
    \left( I_{z}\cos\zeta - I_{y}\sin\zeta \right)
  \right],     
  \label{eq:u2}
\end{align}
where $\omega'_{2}$ is the RF profile seen in the nutating frame and $\zeta$ is a phase factor, we obtain the corresponding RF Hamiltonian $\mathcal{H}_{1} + \mathcal{H}_{2}$ \textit{back from} the target propagetor $U$ using $i \dot{U} U^{-1}$\cite{Anderson1962a} to be
\begin{align}
    \omega_{1}I_{x} + \omega'_{2}(t) \left[ I_{z} \cos(\omega_{1}t+\zeta) - I_{y}\sin(\omega_{1}t+\zeta) \right].
    \label{eq:hrf}
\end{align}
With $\zeta=0$, for instance, we extract the amplitude $\omega_{\mathrm{a}}(t)$ and phase $\phi(t)$ of the RF field to be
\begin{align}
    \omega_{\mathrm{a}}(t) &= \left[ \omega_{1}^{2} + (\omega'_{2}(t) \sin\omega_{1}t)^{2} \right]^{\frac{1}{2}}, 
    \label{eq:omega_a} \\
    \phi(t) &= \tan^{-1}\left[ 
      -\frac{\omega'_{2}(t)\sin\omega_{1}t}{\omega_{1}} 
    \right].
    \label{eq:phi}
\end{align}
The $I_{z}$ term in Eq.~(\ref{eq:hrf}) can be implemented either by audio field applied along the static field in the laboratory frame, frequency modulation of the RF field, or additional phase modulation $\phi'(t)$ given by
\begin{align}
    \phi'(t) = -\int dt' \omega'_{2}(t')\cos\omega_{1}t'.
    \label{eq:phiprime}
\end{align}

Figure~\ref{fig:rf-profile}(a) depicts the nutation-frame RF-field strength $\omega'_{2}(t)$ during ADNF followed by ARNF, starting with the initial value $\omega_{2}$, gradually lowering and vanishing at the end of the interval $\tau$, staying null for $T$, and ramping on.
The corresponding rotating-frame RF amplitude $\omega_{a}(t)$ is schematically described in Fig.~\ref{fig:rf-profile}(b).
In this work, we adopt a smooth and analytically integrable function for $\omega'_{2}(t)$\cite{supp}.
Figure~\ref{fig:rf-profile}(c)-(f) show how the amplitude $\omega_{\mathrm{a}}$ and the phase $\phi+\phi'$ of the RF signal change with time 
for $\omega_{1}=2\pi\cdot 20$~kHz, $\omega_{2}=2\pi\cdot 20$~kHz, and $\tau=2$~ms.
In Fig.~\ref{fig:rf-profile}(d) and (f) plotted are, in addition to the magnified view of the data shown in (c) and (e), experimentally measured amplitude and phase of the modulated RF signals generated with an in-house digital NMR spectrometer\cite{Takeda2007,Takeda2008,Takeda2011a}, confirming that the profiles of the generated signals were as intended.

To ensure that the system is initially in an eigenstate of $\mathcal{H}_{2}^{\mathrm{nut}}$, we may set $\zeta=0$, so that $\bm{B}_{2}(t=0)$ is parallel to the laboratory-frame and rotating-frame $z$-axis.
Since the equilibrium magnetization also aligns along $\bm{B}_{0} \parallel z$, the magnetization is automatically locked to the initial $\bm{B}_{2}$ without a $\pi/2$ pulse\cite{Wang2023}.
That is, the thermal equilibrium state is ensured to be an operational, initialized state ready for ADNF.
Another choice of the direction, e.g., $\bm{B}_{2} \parallel y$, is also possible, but requires an initial $\pi/2$ pulse flipping the magnetization to the direction of $\bm{B}_{2}$.


\begin{figure}[htbp]
    \centering
    \includegraphics[width=0.95\linewidth]{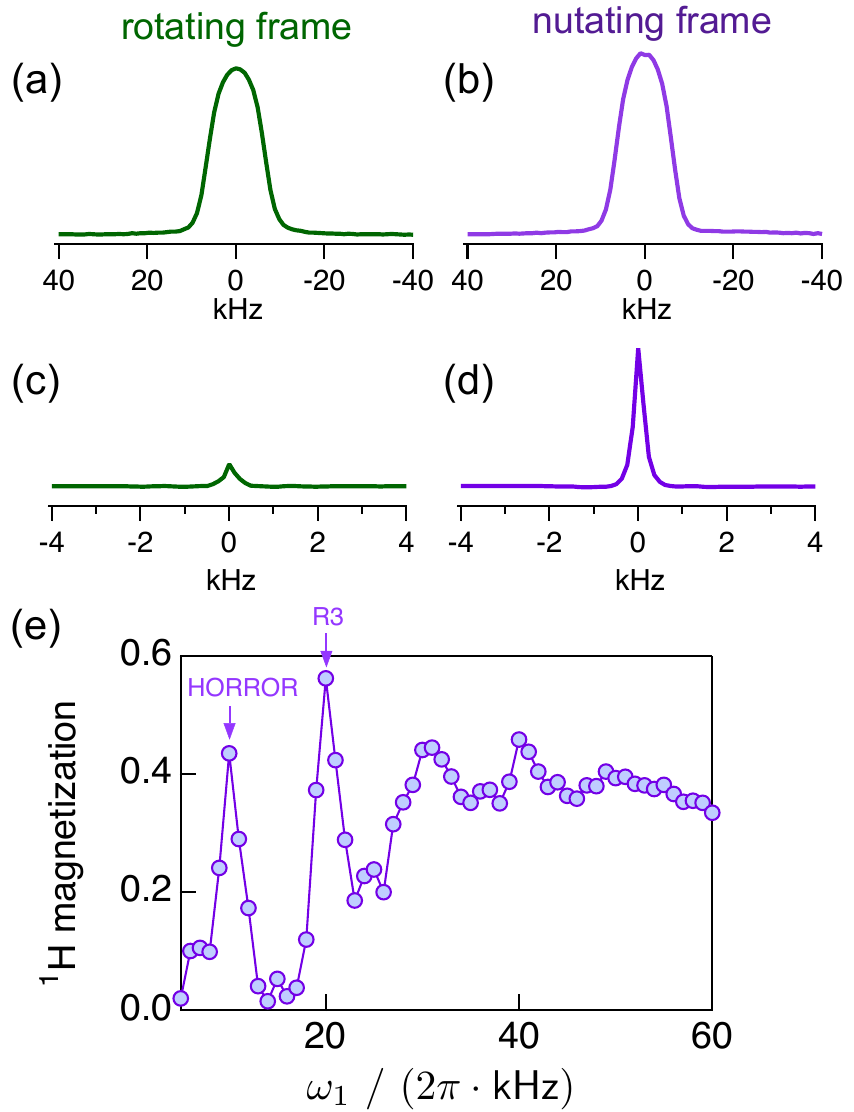}
    \caption{\label{fig:adrf-adnf-comparison} $^{1}$H NMR spectra of adamantane under (a)(b) static ($\omega_{\mathrm{r}}=0$) condition and (c)(d) under MAS at $\omega_{\mathrm{r}}=2\pi\cdot 20$~kHz in 9.4~T. The spectra were obtained by Fourier transforming $^{1}$H free induction decays acquired after implementing (a)(c) ADRF-ARRF (b)(d) ADNF-ARNF sequences ($\omega_{1}=\omega_{\mathrm{r}}=2\pi\cdot 20$~kHz). (e) $\omega_{1}$ dependence of the $^{1}$H magnetization normalized by that in thermal equilibrium.}
\end{figure}

Figure~\ref{fig:adrf-adnf-comparison} compares creation of $^{1}$H dipolar order followed by conversion back to the observable magnetization in the rotating and the nutating frames obtained in a stationary powder of adamantane (C$_{10}$H$_{16}$) and in the same sample under MAS at a spinning rate $\omega_{\mathrm{r}}$ of $2\pi \cdot 20$~kHz. 
In the static sample, dipolar-order creation and conversion worked well for both ADRF-ARRF and ADNF-ARNF (Fig.~\ref{fig:adrf-adnf-comparison}(a)(b)), showing recovery of the $^{1}$H magnetization from the dipolar order.
Conversely, under MAS, the conventional, rotating-frame version resulted in poor conversion efficiency (Fig.~\ref{fig:adrf-adnf-comparison}(c)), which is in line with previous observations and ascribed to that the rotating-frame dipolar interaction $\mathcal{H}_{\mathrm{D}}^{\mathrm{rot}}$ 
written as\cite{Maricq1979,Mehring1983}
\begin{align}
    \mathcal{H}_{\mathrm{D}}^{\mathrm{rot}} &= D(t) \left[3 I_{1z} I_{2z} - \bm{I}_{1}\cdot\bm{I}_{2} \right], \label{eq:hd} \\
    D(t) &= G_{1} \cos(\gamma_{\mathrm{D}} + \omega_{\mathrm{r}}t )
           + G_{2} \cos(2\gamma_{\mathrm{D}} + 2\omega_{\mathrm{r}}t ), 
\label{eq:dt}
\end{align} 
is sinusoidally modulated at frequencies $\omega_{\mathrm{r}}$ and $2\omega_{\mathrm{r}}$ and averaged out in a time scale of the order of $\omega_{\mathrm{r}}^{-1}$.
Here, $G_{1}=-\frac{\sqrt{2}}{4}d\sin 2 \beta_{\mathrm{D}}$ and $G_{2}=\frac{1}{4}d \sin^{2}\beta_{\mathrm{D}}$, where $d=-(\mu_{0}/4\pi) \gamma^{2}\hbar r^{-3}$ is the dipolar coupling constant, $r$ is the distance between the spin pair, and $(\alpha_{\mathrm{D}}=0, \beta_{\mathrm{D}}, \gamma_{\mathrm{D}})$ are the spatial Euler angles connecting the orientation of the axially symmetric tensor of the dipolar interaction in a coordinate system fixed in the spinning sample container and should not be confused with the spin-space transformation from the laboratory to the rotating, and from the rotating to the nutating frames.

Strikingly, the current, nutating-frame implementation of demagnetization/remagnetization resulted in as efficient retention of the population as in the case of stationary sample (Fig.~\ref{fig:adrf-adnf-comparison}(d)).
Here, the nutation rate $\omega_{1}$ was set to satisfy rotary resonance recoupling (R3) condition\cite{Oas1988}, i.e., $\omega_{1}=\omega_{\mathrm{r}}=2\pi\cdot 20$~kHz.
As demonstrated in Fig.~\ref{fig:adrf-adnf-comparison}(e), efficient creation of nutating-frame dipolar order under MAS requires dipolar recoupling, such as R3\cite{Oas1988} and HORROR\cite{Nielsen1994} ($\omega_{1}=\omega_{\mathrm{r}}/2$), which leaves a \textit{finite}, first-order average Hamiltonian of dipolar interactions among the $^{1}$H spins.

\begin{figure}[htbp]
    \centering
    \includegraphics[width=0.85\linewidth]{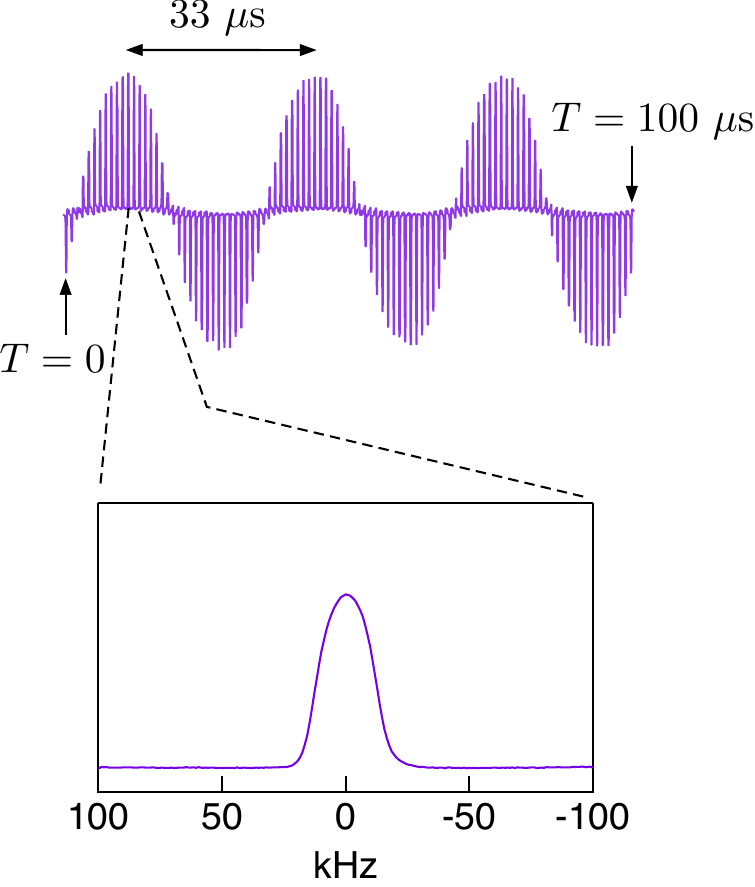}
    \caption{\label{fig:T-dep} Concatenated plots of $^{1}$H spectra of adamantane acquired with varied time intervals $T$ between ADNF and ARNF operations. The nutation rate $\omega_{1}$ was set to $2\pi\cdot 30$~kHz, and $\omega_{2}$ was $2\pi\cdot 18$~kHz.}
\end{figure}

One potential argument against nutating-frame dipolar-order is that the magnetization may somehow have been locked along the first nutation field $\bm{B}_{1}$.
To rule this out and prove that the magnetization has indeed undergone transformation to the nutating-frame dipolar order and then restored back into the Zeeman order, we examined $^{1}$H spectra for stepwisely incremented retention-time $T$ with such a phase cycle that cancels out the spin-locked magnetization in the rotating frame\cite{supp}.
Figure~\ref{fig:T-dep} show a concatenated plot of experimental $^{1}$H spectra. 
The observed periodic profile indicates creation of dipolar order in the nutating frame, which initially coincides with the rotating frame but changes its relative orientation at a rate $\omega_{1}$ around the latter's $x$-axis.
After successful dipolar-order creation and Zeeman-order restoration, the $^{1}$H magnetization is expected to orient along the axis of the \textit{nutating-frame}.
Thus, the restored magnetization at the end of the sequence ought to have gained rotation by an angle $\omega_{1}(2\tau+T)$ around the rotating-frame $x$-axis.
Since it is only the magnetization component in the $xy$ plane of the rotating frame that contributes to the observable nuclear induction signal, the periodic behavior in Fig.~\ref{fig:T-dep} confirms that dipolar order has indeed been created \textit{in the nutating frame}.
To compensate such an oscillatory behavior, we can choose the phase $\zeta$ in the ARNF process (Eq.~(\ref{eq:u2})) in such a way that the recovered magnetization is ensured to be in the $xy$-plane of the rotating-frame.
Indeed, we did so in the ADNF-ARNF experiments shown in Fig.~\ref{fig:adrf-adnf-comparison} and in the following measurements.

\begin{figure}[htbp]
    \centering
    \includegraphics[width=0.95\linewidth]{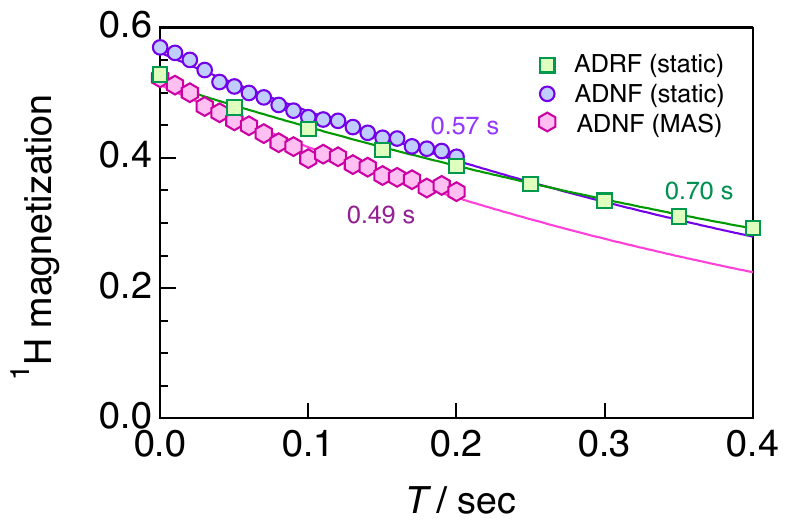}
    \caption{\label{fig:decay} Dependence of the time interval ($T$) between demagnetization and remagnetization operations on the resultant $^{1}$H magnetization of adamantane obtained with ADRF-ARRF in the static condition (squares), ADNF-ARNF in the static condition (circles), and ADNF-ARNF under MAS (hexagons). Solid lines represent curve fitting with an exponentially decaying function. The decay time constants are also indicated.}
\end{figure}

Figure~\ref{fig:decay} shows experimental dependence of the restored magnetization on the retention time ($T$) for ADRF, static ADNF, and ADNF under MAS.
The data measuring the dipolar-order's lifetime decayed exponentially with time constants of 0.70~s, 0.57~s, and 0.49~s, respectively.
The experimental observation that the nutating-frame dipolar order, whether in the static condition or under MAS, lasted over the same order of duration as the rotating-frame dipolar order is not trivial but can be understood by considering that, for an adiabatically changing Hamiltonian, its eignestate at one moment continues to be the eigenstate of the instantaneous, time-dependent Hamiltonian.
In the present context, the magnetization, or Zeeman order, initially being an eigenstate of the RF Zeeman interaction in either the rotating or the nutating frame, eventually settles into another eigenstate of the time-averaged secular part of the dipolar interaction $\overline{\mathcal{H}}_{\mathrm{D}}$.
Smooth transfer from the Zeeman order to the nutating-frame dipolar order is secured by the adiabatic nature of operation.

For static ADNF with $\bm{B}_{1} \parallel x$, $\overline{\mathcal{H}}_{\mathrm{D}}$ is given by\cite{Rhim1970,Rhim1971,Rhim1971a}
\begin{align}
    \overline{\mathcal{H}}_{\mathrm{D}}
    = - \frac{1}{2} \exp\left(-i\frac{\pi}{2}I_{y}\right) \mathcal{H}_{\mathrm{D}}^{\mathrm{rot}} \exp\left(i\frac{\pi}{2}I_{y}\right).
\end{align}
This is quite similar to the rotating-frame dipolar interaction except for the factor of only $-1/2$ and change of the basis set represented by the unitary operator $e^{-i(\pi/2)I_{y}}$.
Thus, the dynamics and thereby the time scale of the relaxation processes would be close to the case of ADRF.
Conversely, the secular dipolar interaction for ADNF under MAS with the HORROR ($k=1$) and R3 ($k=2$) recoupling conditions, expressed as\cite{supp}
\begin{align}
  \overline{\mathcal{H}}_{\mathrm{D}} =
  \frac{3}{4}G_{k}\left[ 
   \cos k \gamma (I_{x} S_{x} - I_{y} S_{y}) - \sin k \gamma (I_{y} S_{x} + I_{x}S_{y})
  \right],
\end{align}
is also represented in a matrix form as
\begin{align}
    \overline{\mathcal{H}}_{\mathrm{D}} =
    \frac{3}{8}G_{k}
    \left[
    \begin{array}{cccc}
        0 & 0 & 0 & e^{ik\gamma_{\mathrm{D}}} \\
        0 & 0 & 0 & 0 \\
        0 & 0 & 0 & 0 \\
        e^{-ik\gamma_{\mathrm{D}}} & 0 & 0 & 0 
    \end{array}
    \right].
\end{align}
Thus, the first-order nutating-frame dipolar order under MAS is the population occupied to the eigenstate of the double-quantum interaction.
The intriguing observation that the decay time constant was down only 14~\% compared to the static case indicates that the relaxation of dipolar order is insensitive to the structure of the secular component of the Hamiltonian.

To conclude, it is possible to create dipolar order under MAS in the nutating frame through adiabatic demagnetization of the nutating-frame spin-locking field $\bm{B}_{2}$, likewise the well established dipolar order created in the laboratory and the rotating frames.
Dipolar order in the nutating frame can be made robust under MAS by adjusting the nutating field $\bm{B}_{1}$ to meet one of the dipolar recoupling conditions.
Just like transformation from the laboratory frame to the rotating frame eliminates the static field $\bm{B}_{0}$ in the latter, the nutating field $\bm{B}_{1}$ vanishes through further transformation into the nutating frame, where the spins' interactions acquire time dependence and are also subject to manipulation by the remaining field $\bm{B}_{2}$, which can be designed through reverse engineering of the RF Hamiltonian and implemented by amplitude/phase modulation of the RF field applied to the spin system.
The nutating frame provides another arena in which nuclear spins show rich dynamics, an example of which is dipolar order creation as demonstrated here.
Albeit being unobservable directly, dipolar order is reversibly converted to/from an observable magnetization and thus retains well-defined spin temperature. Through thermal contact, dipolar order serves not only for probing coupling with the fluctuating environment but also for cooling down another isotope's spins. Therefore, expanding the arena in which dipolar order comes into play stands out as a groundbreaking example of quantum control.

We are grateful to Prof.~Matthias Ernst for fruitful discussions.
This work has been supported by the MEXT Quantum Leap Flagship Program (MEXT Q-LEAP) (grant number JPMXS0120330644).

\bibliography{adnf}

\clearpage 
\onecolumngrid
\appendix
\renewcommand{\thefigure}{A\arabic{figure}}
\renewcommand{\theequation}{A\arabic{equation}}
\setcounter{figure}{0}
\setcounter{equation}{0}

\section{Adiabatic demagnetization --- rotating vs nutating frames}
\begin{figure*}[htbp]
    \centering
    \includegraphics[width=\linewidth]{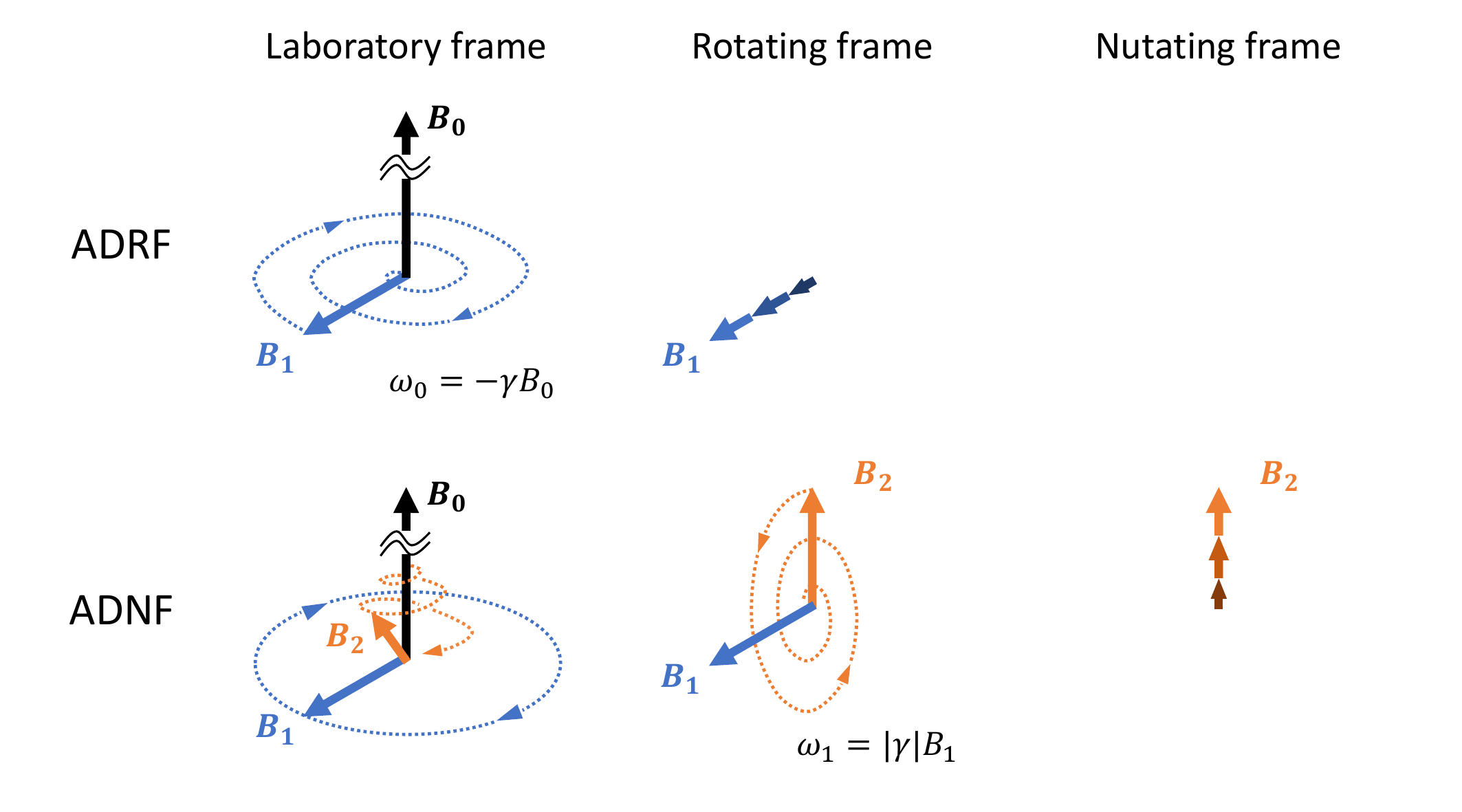}
    \caption{\label{fig:concept} A static magnetic field $\bm{B}_{0}$ and RF fields $\bm{B}_{1}$ and $\bm{B}_{2}$, and their trajectories during ADRF (top) and ADNF (bottom).}
\end{figure*}

Figure~\ref{fig:concept} shows schematic drawings of the magnetic-field components $\bm{B}_{0}$, $\bm{B}_{1}$, and $\bm{B}_{2}$, together with their trajectories during implementing ADRF (Fig.~\ref{fig:concept}, top) and ADNF (Fig.~\ref{fig:concept}, bottom).
The RF magnetic field $\bm{B}_{1}$ of ADRF seen in the laboratory frame rotates about the static field $\bm{B}_{0}$ at the carrier frequency $\omega_{0}$ set at the Larmor precession frequency $-\gamma B_{0}$.
Meanwhile, the amplitude of $\bm{B}_{1}$ is gradually reduced and eliminated.
As a result, the trajectory of $\bm{B}_{1}$ follows a spiral path.
In the rotating frame, the static field $\bm{B}_{0}$ is invisible and $\bm{B}_{1}$ acts as an adiabatically demagnetizing field, reducing its length without changing its direction.
Note that, for explanatory clarity, the spiral line is schematically drawn much sparsely than the realistic trajectories.
In ADNF, (Fig.~\ref{fig:concept}, bottom), the magnitude of $\bm{B}_{1}$ is kept fixed, so that it follows a circular path in the laboratory frame and stay still in the rotating frame.
In addition applied is $\bm{B}_{2}$ designed to draw a spiral path in the nutating frame (Fig.~\ref{fig:concept}, bottom, middle).
Again, the drawing is schematic, so the actual trajectory is a more dense spiral.
The path of $\bm{B}_{2}$, being somewhat complicated in the laboratory frame, becomes a demagnetizing field in the nutating frame.

\section{\label{sec:rfprofiles}RF field profiles for ADNF}
\begin{figure*}[htbp]
    \centering
    \includegraphics[width=0.8\linewidth]{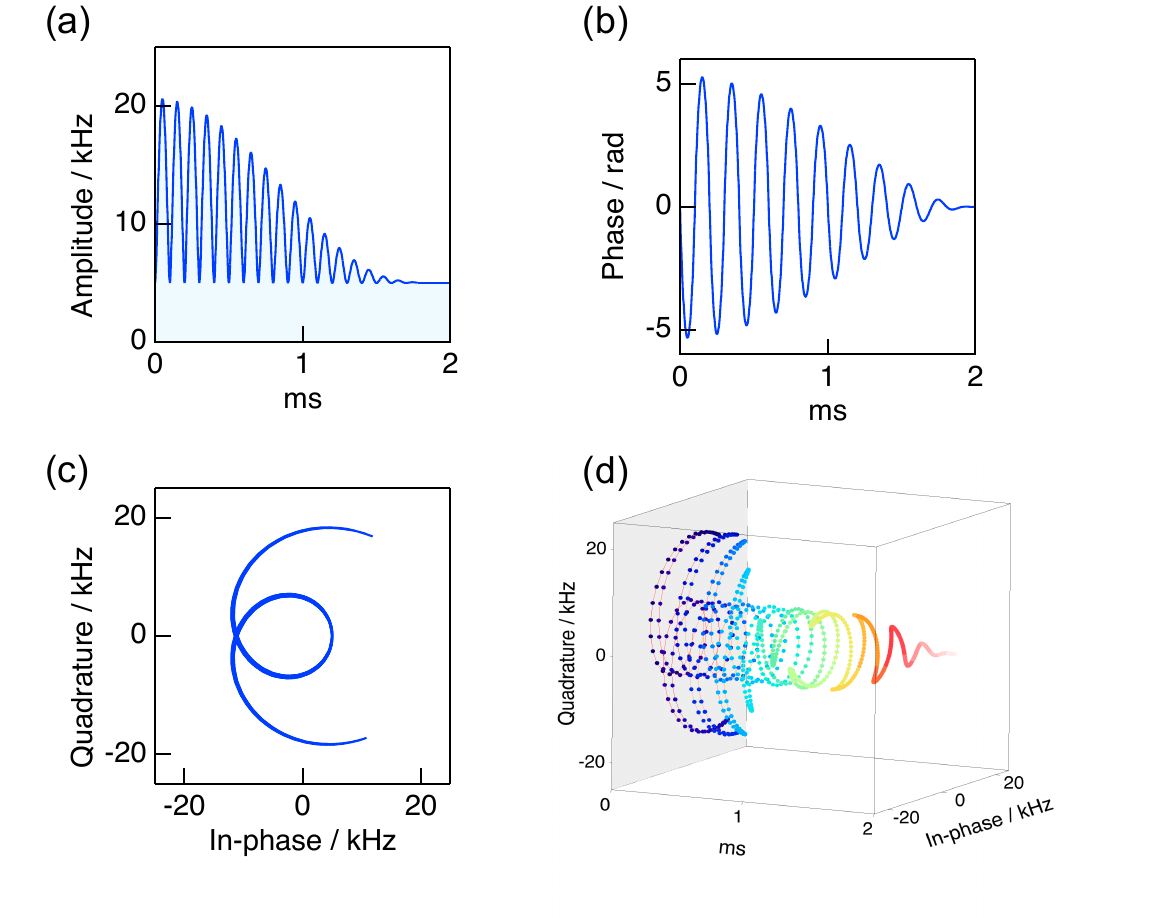}
    \caption{\label{fig:5-20} Time dependence of (a) amplitude and (b) phase of an RF pulse implementing ADNF with $\omega_{1}=2\pi\cdot 5$~kHz, $\omega_{2}=2\pi \cdot 20$~kHz, and duration $\tau=2$~ms. (c) A parametric plot of the in-phase and quadrature components of the RF field of the same pulse. (d) a three-dimensional plot visualizing the time dependence of the parametric plot.}
\end{figure*}

\begin{figure*}[htbp]
    \centering
    \includegraphics[width=0.8\linewidth]{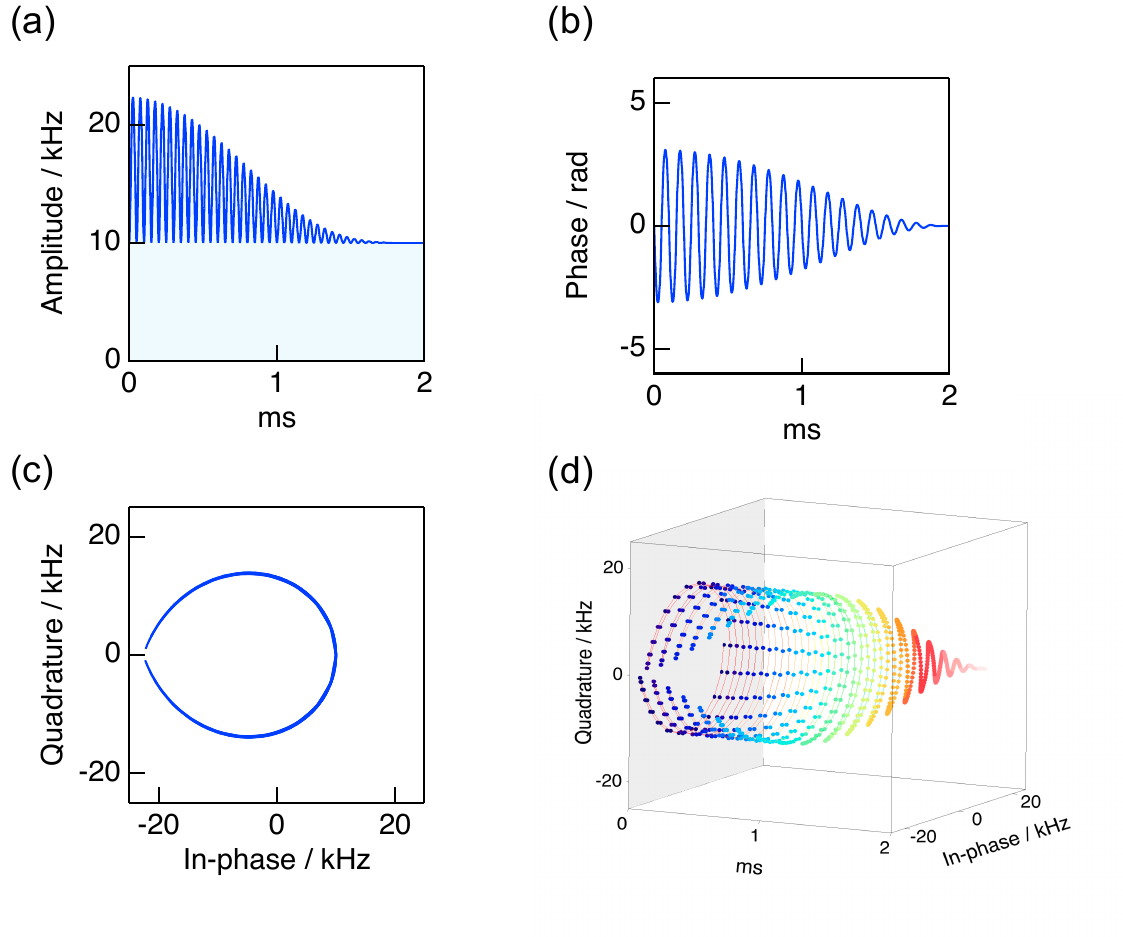}
    \caption{\label{fig:10-20} Time dependence of (a) amplitude and (b) phase of an RF pulse implementing ADNF with $\omega_{1}=2\pi\cdot 10$~kHz, $\omega_{2}=2\pi \cdot 20$~kHz, and duration $\tau=2$~ms. (c) A parametric plot of the in-phase and quadrature components of the RF field of the same pulse. (d) a three-dimensional plot visualizing the time dependence of the parametric plot.}
\end{figure*}
For the profile $\omega'_{2}(t)$ of the RF field of ADNF seen in the nutating frame starting with the initial value $\omega_{2}$ at $t=0$ and ending with zero at $t=\tau$,
we adopt a smooth and analytically integrable function
\begin{align}
  \omega'_{2}(t) = \frac{\omega_{2}}{2} \left[ 1+ \cos\left( \frac{\pi t}{\tau} \right) \right], \quad 0 \le t \le \tau.
\end{align}
Straightforward calculation of Eq.~(5) leads to the additional phase modulation $\phi'(t)$ written by
\begin{align}
    \phi'(t) = -\frac{\omega_{2}t}{2}\mathrm{sinc}(\omega_{1}t) 
      - \frac{\omega_{2}t}{4} \mathrm{sinc} \left[\left( \frac{\pi}{\tau} + \omega_{1} \right) t \right]
      - \frac{\omega_{2}t}{4} \mathrm{sinc} \left[\left( \frac{\pi}{\tau} - \omega_{1} \right) t \right].
    \label{eq:phi_prime}
\end{align}
Figure~\ref{fig:5-20}(a)(b) show an example of RF amplitude modulation $\omega_{\mathrm{a}}(t)$ of $2\pi \cdot 5$~kHz, the initial nutating-frame RF-field strength $\omega_{2}$ of $2\pi\cdot 20$~kHz, and the ADNF interval $\tau$ of 2~ms.
Figure~\ref{fig:5-20}(c)(d) are parametric plots of the in-phase and quadrature components of this RF field.
Figure~\ref{fig:10-20} shows another example with $\omega_{1}=2\pi\cdot 10$~kHz, $\omega_{2}=2\pi\cdot 20$~kHz, and $\tau=2$~ms.

For ARNF implemented from $t=\tau+T$ to $2\tau+T$, we set a monotonically \textit{increasing} function starting from 0 at $t=\tau+T$ and end with $\omega_{2}$ at $t=2\tau+T$ as
\begin{align}
    \omega'_{2}(t) = \frac{\omega_{2}}{2} \left[
       1-\cos\left[
            \frac{\pi}{\tau}(t-\tau-T)
        \right]
    \right].
\end{align}

\section{signal averaging with phase cycling}
To obtain the $^{1}$H spectra shown in Fig.~3 and Fig.~4, we accumulated the acquired $^{1}$H signals over 8 times for each of the data taken with stepwisely incremented values of $T$.
In the ADNF part, i.e., $0<t<\tau$ (see Fig.~1), the phase factor $\zeta$ (introduced in Eq.~(1)) that determines the initial phase of the second nutating field $\bm{B}_{2}$ around the first, $\bm{B}_{1}$, was kept to be 0 throughout the experiment, while the value of $\zeta$ for the ARNF part $(\tau+T < t < 2\tau+T)$ was set alternately to 0 and $\pi$.
That is, $\zeta=0$ for the odd-times signal accumutations and $\zeta=\pi$ for the even-times.
In addition, the acquired data of even-numbered times were subtracted from thouse of the odd-numbered times.
Since the magnetization components, if any, locked along the rotating-frame spin locking field $\bm{B}_{1}$ is not affected by the cycle of the phase factor $\zeta$, they cancel out by the alternate subtraction.
Conversely, the magnetizations having undergone transformation to the nutating-frame dipolar order and back to the magnetization are inverted in the even-numbered shots with $\zeta=\pi$, and add up with subtraction.

\section{Derivation of Eqs.~(9)}
The nutating-frame dipolar interaction $\mathcal{H}_{\mathrm{D}}^{\mathrm{nut}}$ is given by transformation of the rotating-frame dipolar interaction $\mathcal{H}_{\mathrm{D}}^{\mathrm{rot}}$ (Eq.~(6)) into the interaction frame with respect to the RF Zeeman interaction $\mathcal{H}_{1}$.
\begin{align}
    \mathcal{H}_{\mathrm{D}}^{\mathrm{nut}}
    &= U_{1}^{-1} \mathcal{H}_{\mathrm{D}}^{\mathrm{rot}} U_{1},
\end{align}
where $U_{1} = \exp(-i \mathcal{H}_{1}t) = \exp(-i\omega_{1}t I_{x})$.
Since $\bm{I}_{1}\cdot \bm{I}_{2}$ in Eq.~(6) is invariant under spin-space rotation, we shall take a look at how the $I_{1z}I_{2z}$ term changes:
\begin{align}
    I_{1z} I_{2z}
    & \rightarrow \exp(i\omega_{1}t I_{x}) I_{1z} I_{2z} \exp(-i\omega_{1}t I_{x}) \\
    &= \left( I_{1z} \cos\omega_{1}t + I_{1y}\sin\omega_{1}t \right)
    \left( I_{2z} \cos\omega_{1}t + I_{2y}\sin\omega_{1}t \right) \\
    &= \frac{1}{2} \left[ 
    (1+\cos 2\omega_{1}t)I_{1z}I_{2z} + (1-\cos 2\omega_{1}t) I_{1y}I_{2y} \right. \nonumber \\
    & \qquad \left. \sin 2\omega_{1}t I_{1y}I_{2z} + \sin 2\omega_{1}t I_{1z}I_{2y}
    \right].
\end{align}
Then, we toggle the system around $y$ by $\pi/2$ as we do in Eq.~(8), obtaining
\begin{align}
    \frac{1}{2} \left[ 
    (1+\cos 2\omega_{1}t)I_{1x}I_{2x} + (1-\cos 2\omega_{1}t) I_{1y}I_{2y} + \sin 2\omega_{1}t I_{1y}I_{2x} + \sin 2\omega_{1}t I_{1x}I_{2y}
    \right].
\end{align}
When we multiply this by $D(t)$ (Eq.~(7)) given by
\begin{align}
D(t) = G_{1} \cos(\gamma_{\mathrm{D}} + \omega_{\mathrm{r}}t )
           + G_{2} \cos(2\gamma_{\mathrm{D}} + 2\omega_{\mathrm{r}}t ),    
\end{align}
a finite (non-zero) time average $\overline{\mathcal{H}}_{\mathrm{D}}$ results when $\omega_{1}=\omega_{\mathrm{r}}$ (HORROR) or $\omega_{1}=2\omega_{\mathrm{r}}$ (R3).
In the case of HORROR,
\begin{align}
    \overline{\mathcal{H}}_{\mathrm{D}} = \frac{3}{4}G_{1} \left[
        \cos \gamma_{\mathrm{D}} (I_{1x}I_{2x} - I_{1y}I_{2y})
      - \sin \gamma_{\mathrm{D}} (I_{1y}I_{2x} + I_{1x}I_{2y}) 
     \right],
\end{align}
whereas for R3,
\begin{align}
    \overline{\mathcal{H}}_{\mathrm{D}} = \frac{3}{4}G_{2} \left[
        \cos 2\gamma_{\mathrm{D}} (I_{1x}I_{2x} - I_{1y}I_{2y})
      - \sin 2\gamma_{\mathrm{D}} (I_{1y}I_{2x} + I_{1x}I_{2y}) 
     \right],
\end{align}
which can be summarized in a single formula as described in Eq.~(9) with $k=1,2$.

\end{document}